\documentclass[final,1p]{elsarticle}
\biboptions{square}
\usepackage{amsmath}
\usepackage{amssymb}
\usepackage{graphicx}
\usepackage{indentfirst}

\DeclareMathSymbol{\lang}{\mathord}{symbols}{"68}
\DeclareMathSymbol{\rang}{\mathord}{symbols}{"69}
\DeclareMathSymbol{\openbra}{\mathord}{symbols}{"68}
\DeclareMathSymbol{\closeket}{\mathord}{symbols}{"69}
\DeclareMathOperator{\Rre}{Re}
\DeclareMathOperator{\Iim}{Im}

\newcommand{\ket}[1]{{| #1 \closeket}}

\renewcommand{\phi}{\varphi}
\renewcommand{\epsilon}{\varepsilon}
\begin{document}
\begin{frontmatter}

\title{Nonadiabatic quantum chaos in atom optics}
\author{S.V.~Prants}
\ead{prants@poi.dvo.ru, tel.007-4232-312602, fax 007-4232-312573}
\ead[url]{http://dynalab.poi.dvo.ru}
\address{Laboratory of Nonlinear Dynamical Systems,
Pacific Oceanological Institute of the Russian Academy of Sciences, 
Baltiiskaya St., 43, 690041 Vladivostok, Russia}
\begin{abstract}
Coherent dynamics of atomic matter waves in a standing-wave 
laser field is studied. In the dressed-state picture, wave packets of 
ballistic two-level atoms propagate simultaneously in two optical potentials. 
The probability to make a transition from one potential to another one 
is maximal when centroids of wave packets cross the field nodes and 
is given by a simple formula with the single exponent, the Landau--Zener 
parameter $\kappa$. If $\kappa \gg 1$, the motion is essentially adiabatic.
If $\kappa \ll 1$, it is (almost) resonant and periodic. If  
$\kappa \simeq 1$, atom makes  nonadiabatic transitions with a splitting of 
its wave packet at each node and strong complexification of the wave function 
as compared to the two other cases. This effect is referred as nonadiabatic 
quantum chaos. Proliferation of wave packets at $\kappa \simeq 1$ is shown to be 
connected closely with chaotic center-of-mass motion in the semiclassical 
theory of point-like atoms with positive values of 
the maximal Lyapunov exponent. 
The quantum-classical correspondence established is justified by the fact 
that the Landau--Zener parameter $\kappa$ specifies the regime of the 
semiclassical dynamical chaos in the map simulating chaotic center-of-mass 
motion. Manifestations of nonadiabatic quantum chaos are found in the 
behavior of the momentum and position probabilities. 
\end{abstract}
\end{frontmatter}


\section{Introduction}
The mechanical action of light upon neutral atoms has been comprehensively 
studied since the pioneer works of Lebedev, Gerlah and Stern, Kapitza and 
Dirac and Frisch. 
The light pressure force provides optical cooling and trapping of atoms 
\cite{Letokhov}. In the last two decades, cold atoms in standing-wave optical 
fields have been used to study quantum chaos. The proposal 
\cite{GSZ92} to study atomic dynamics in 
a far-detuned modulated standing wave made atomic optics  
a testing ground for quantum chaos. A number of impressive experiments have 
been carried out in accordance with this proposal \cite{Raizen,Steck01,HH01}. 
New possibilities are opened if one works near the atom-field resonance where 
the interaction between the internal and external atomic degrees of freedom is 
intense \cite{JETPL01,PRA01,JETPL02}.

Dynamical chaos in classical mechanics is a special kind of random-like motion 
without any noise and/or random parameters. It is characterized by exponential 
sensitivity of trajectories in the phase space to small variations in 
initial conditions and/or control parameters. Such sensitivity does not 
exist in isolated quantum systems because their evolution is unitary, and there is 
no well-defined notion of a quantum trajectory. Thus, there is a fundamental 
problem of emergence of classical dynamical chaos from more profound 
quantum mechanics which is known as quantum chaos problem and the related 
problem of quantum-classical correspondence. 
In a more general context it is a problem of wave chaos. It is clear now that 
quantum chaos, microwave, optical, and acoustic chaos \cite{Haake,Reich,Stockmann,
Makarov} have much in common. The common practice is to construct an analogue 
for a given wave object 
in a semiclassical (ray) approximation and study its chaotic properties (if any) by 
well-known methods of dynamical system theory. Then, it is necessary to solve 
the corresponding linear wave equation in order to find manifestations of 
classical chaos in the wave-field evolution in the same range of the control 
parameters. If one succeeds in that the quantum-classical or the wave-ray 
correspondence are announced to be established. 

In atom optics \cite{K90} one quantizes both the atomic internal and 
translational degrees of freedom. The atom is treated as a wave
packet which undergoes deformations in the process of exchange of energy and 
momentum quanta with a light wave. 
Quantization of the translation motion provides an entanglement of the 
internal and external degrees of freedom. Any changes in the form of
the wave packet will affect the internal state of the atom and vice versa 
\cite{JETP09,JETPL10}.
The optical field provides a tool to manipulate the atomic matter waves. 
In atom optics the Schr\"odinger equation for the probability amplitudes 
constitutes  a linear infinite-dimensional dynamical system which is  governed 
by an external force if the field is treated as a classical wave. 

In the semiclassical approximation, atom with quantized internal dynamics is  
treated as a point-like particle with the Hamilton--Schr\"odinger 
equations of motion constituting a nonlinear dynamical system. 
A number  of nonlinear Hamiltonian and dissipative dynamical effects have been
found with such a system including chaotic Rabi oscillations, chaotic atomic 
transport, dynamical fractals, synchronization, chaotic walking, and 
L\'evy flights \cite{PRE02,JETP03,PRA05,PRA07,PRA08,EPL08}. 
Similar and new effects 
have been found numerically and described analytically with two-level atoms 
in a losseless cavity with a single quantized mode in the framework of the 
Jaynes-Cummings model \cite{PLA03,CNSNS03,PRA06,CNSNS07} and the 
Tavis-Cummings model \cite{Chot}. It has been shown that the coupled 
atom-field dynamics in a cavity can be unstable under appropriate conditions 
in the absence of any kind of interaction with environment. This kind of 
quantum instability manifests itself in fractal chaotic scattering of atoms 
\cite{PLA03,CNSNS03,PRA06,CNSNS07}, in strong variations of reduced quantum 
purity and entropy \cite{PRA06,CNSNS07,Chot}, correlating with the respective 
maximal Lyapunov exponent, and in exponential sensitivity of fidelity of 
quantum states to small variations of the detuning \cite{PRA06,CNSNS07}.  
 
The main aim of this  paper is to establish a kind of the quantum-classical
correspondence in transport properties of point-like atoms and atomic matter waves 
moving in a standing-waves field. Is the 
coherent evolution of the atomic matter waves really complicated
in that range of the control parameters where the corresponding center-of-mass 
motion has been shown to be chaotic?

\section{Wave-packet motion in a standing light wave}

The Hamiltonian of a two-level atom, moving 
along a one-dimensional classical standing-wave laser field, 
can be written in the frame rotating with the laser frequency $\omega_f$ as follows:
\begin{equation}
\hat H=\frac{\hat P^2}{2m_a}+\frac{\hbar}{2}(\omega_a-\omega_f)\hat\sigma_z-
\hbar \Omega\left(\hat\sigma_-+\hat\sigma_+\right)\cos{k_f \hat X},
\label{Ham}
\end{equation}
where $\hat\sigma_{\pm, z}$ are the Pauli operators for the internal atomic degrees
of freedom, $\hat X$ and $\hat P$ are the atomic position and momentum operators, 
$\omega_a$ and $\Omega$ are the atomic transition and Rabi frequencies, respectively. 
We will work in the momentum representation and expand  
the state vector as  follows:
\begin{equation}
\ket{\Psi(t)}=\int [a(P,t)\ket{2}+b(P,t)\ket{1}]\ket{P}dP,
\label{wavef}
\end{equation}
where $a(P,t)$ and $b(P,t)$ are the probability amplitudes to find atom
at time $t$ with the momentum $P$ in the excited, $\ket{2}$, and ground, 
$\ket{1}$, states, respectively. After some algebra 
one gets the normalized Schr\"odinger equation for the probability amplitudes
\cite{JETP09}
\begin{equation}
\begin{aligned}
i \dot a(p) = \frac12 (\omega_rp^2 - \Delta)a(p) - \frac12[b(p+1) + b(p-1)],
 \\
i \dot b(p) = \frac12 (\omega_rp^2 + \Delta)b(p) - \frac12[a(p+1)+ a(p-1)], 
\end{aligned}
\label{Schodinger}
\end{equation}
where the dot denotes differentiation with respect to dimensionless 
time $\tau \equiv \Omega t$, $p\equiv P/\hbar k_f$, and $x\equiv  k_fX$.
The normalized recoil frequency $\omega_r \equiv
\hbar k_f^2/m_a\Omega$
and the atom-field detuning $\Delta \equiv (\omega_f-\omega_a)/\Omega$ 
are the control parameters.

The probability to find an atom with the momentum $p$ at the moment of time 
$\tau$ is ${\cal P}(p,\tau)= |a(p,\tau)|^2 + |b(p,\tau)|^2$.
The internal atomic state is described by the following 
real-valued combinations of the probability amplitudes:
$u_q(\tau)\equiv 2 \Rre \int dp \left[ a(p,\tau)b^*(p,\tau) \right]$, 
$v_q(\tau)\equiv  -2 \Iim \int dp [a(p,\tau)b^*(p,\tau)]$, 
$z_q(\tau)\equiv \int dp [|a(p,\tau)|^2 - |b(p,\tau)|^2]$,
which are expected values of the synchronized 
(with the laser field) and a quadrature 
components of the atomic electric dipole moment ($u_q$ and $v_q$, respectively) 
and the atomic population inversion, $z_q$.
Varying the value of the Rabi frequency $\Omega$, we can change the value 
of the dimensionless recoil frequency $\omega_r$ with the same atom. 
Working, say, with a cesium atom ($m_a =133$ a.u., 
$\lambda_f = 852.1$ nm, and $\nu_{\rm rec}\simeq2$ KHz), 
we get $\omega_r =10^{-5}$ at $\Omega=100$ MHz. 

We will interpret the wave-packet motion in the dressed-state basis 
\cite{K90,Cohen}
\begin{equation} 
\ket{+}_\Delta=\ket{2}\sin{\Theta}+\ket{1}\cos{\Theta},\ \ket{-}_\Delta=\ket{2}\cos{\Theta}-\ket{1}\sin{\Theta},
\label{dress}
\end{equation}
where $\Theta$ is the mixing angle
\begin{equation} 
\tan{\Theta}\equiv\frac{\Delta}{2\cos{x}} - 
\sqrt{\left( \frac{\Delta}{2\cos{x}} \right)^2 + 1}.
\label{mix}
\end{equation}
These states are eigenstates of an atom 
at rest in a laser field with  the eigenvalues of the quasienergy 
\begin{equation} 
E_\Delta^{(\pm)} = \pm\sqrt{\frac{\Delta^2}{4} +\cos^2{x}}.
\label{energy}
\end{equation}
The probability amplitudes to find the atom 
at point $x$ in those potentials are, respectively 
\begin{equation} 
C_+(x)=a(x)\sin{\Theta}+ b(x)\cos{\Theta},
C_-(x)=a(x)\cos{\Theta}- b(x)\sin{\Theta},
\label{cprob}
\end{equation}
where the amplitudes in the bare-state basis $a(x)$ and  $b(x)$ may be 
computed in the position representation with the help of the Fourier transform
\begin{equation}
a(x)={\rm const}\int_{-\infty}^{\infty}dp'e^{ip'x}a(p'),\ 
b(x)={\rm const}\int_{-\infty}^{\infty}dp'e^{ip'x}b(p').
\label{Four}
\end{equation}

Let us assume that we are able to prepare an atom exactly in one of its dressed 
states, $\ket{+}_\Delta$ or $\ket{-}_\Delta$. Then the atom will move 
in one of the potentials,  $E_\Delta^{(+)}$ or $E_\Delta^{(-)}$, along 
a single trajectory. In quantum mechanics, there is a nonzero probability 
to make a transition to another potential. To estimate this probability  
we write the Hamiltonian of the internal degree of freedom of a two-level 
atom in the basis $\ket{\pm}_{\Delta}$
\begin{equation} 
\hat H_{\rm int}= \hat \sigma_z\cos{x}+\frac{\Delta}{2}\hat\sigma_x.
\label{Hamint}
\end{equation}
Let us linearize the cosine in the vicinity of a node of the standing 
wave and estimate  a small distance the atom makes when crossing the
 node as follows: $\delta x=\omega_r|p_{\rm node}|\tau$ \cite{K90}. 
The quantity $\omega_r|p_{\rm node}|$ 
is a normalized Doppler shift for an atom moving with the momentum 
$|p_{\rm node}|$, i.e., $\omega_D \equiv \omega_r|p_{\rm node}| \equiv k_f
|v_{\rm node}|/\Omega$. The Schr\"odinger
 equation for the probability amplitudes $C_\pm(x)$ in the position  
representation can be written in the form of the second-order equation
\begin{equation} 
\ddot C_+(x)+\left[ i\omega_D  + \frac{\Delta^2}{4} + (\omega_D \tau)^2 
\right] C_+(x) = 0.
\label{cc}
\end{equation}
The asymptotic solution of Eq.~(\ref{cc}),
\begin{equation} 
P_{LZ} = \exp( -\kappa),
\label{LZ}
\end{equation}
gives the probability to make a nonadiabatic or Landau-Zener transition 
from one of the nonresonant potentials to another one specified by  
the Landau--Zener parameter
\begin{equation} 
\kappa \equiv \pi\frac{\Delta^2}{\omega_D}. 
\label{LZparameter}
\end{equation}

There are three regimes of atomic motion.

\begin{enumerate}[1.]
 \item $\kappa\gg 1$. The probability to make the transition is 
exponentially small even when an atom crosses a node. The evolution of the 
atomic wave packet is adiabatic in this case.
 \item $\kappa \ll 1$. The distance between the potentials at the nodes 
is small and the atom changes the potential each time when 
crossing any node with the probability close to unity. In the limit case 
$\Delta=0$,\ the atom moves in the resonant potentials.
 \item $\kappa \simeq 1$. The probability to change the potential 
or to remain in the same one, upon crossing a node, are of the same order. 
In this regime one may expect a proliferation of components of the atomic 
wave packet at the nodes and complexification of the wave function.
\end{enumerate}

\section{Simulation of ballistic wave-packet propagation}

We simulate the evolution of a Gaussian wave packet 
with the variance in the momentum space, $\sigma_p^2=50$, 
$p_0=10^3$, $x_0=0$ and $\omega_r=10^{-5}$. 
The initial average kinetic energy, $\omega_r p^2/2 =5$, is greater than the depth 
of the potential wells, so the atom will move ballistically 
along the positive direction of the standing-wave axis. 

\begin{figure}[htb]\center
\includegraphics[width=0.45\textwidth,clip]{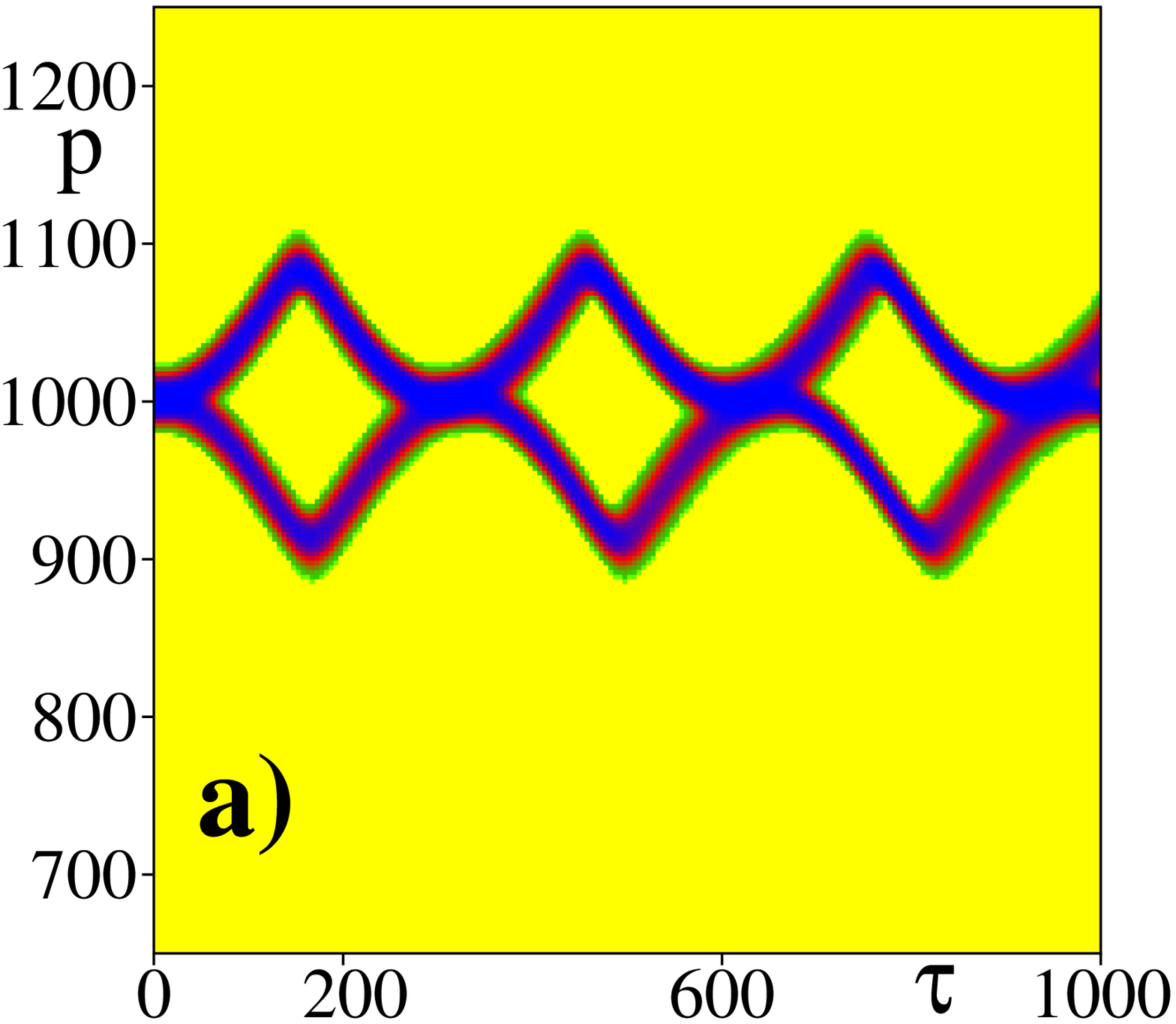}
\includegraphics[width=0.45\textwidth,clip]{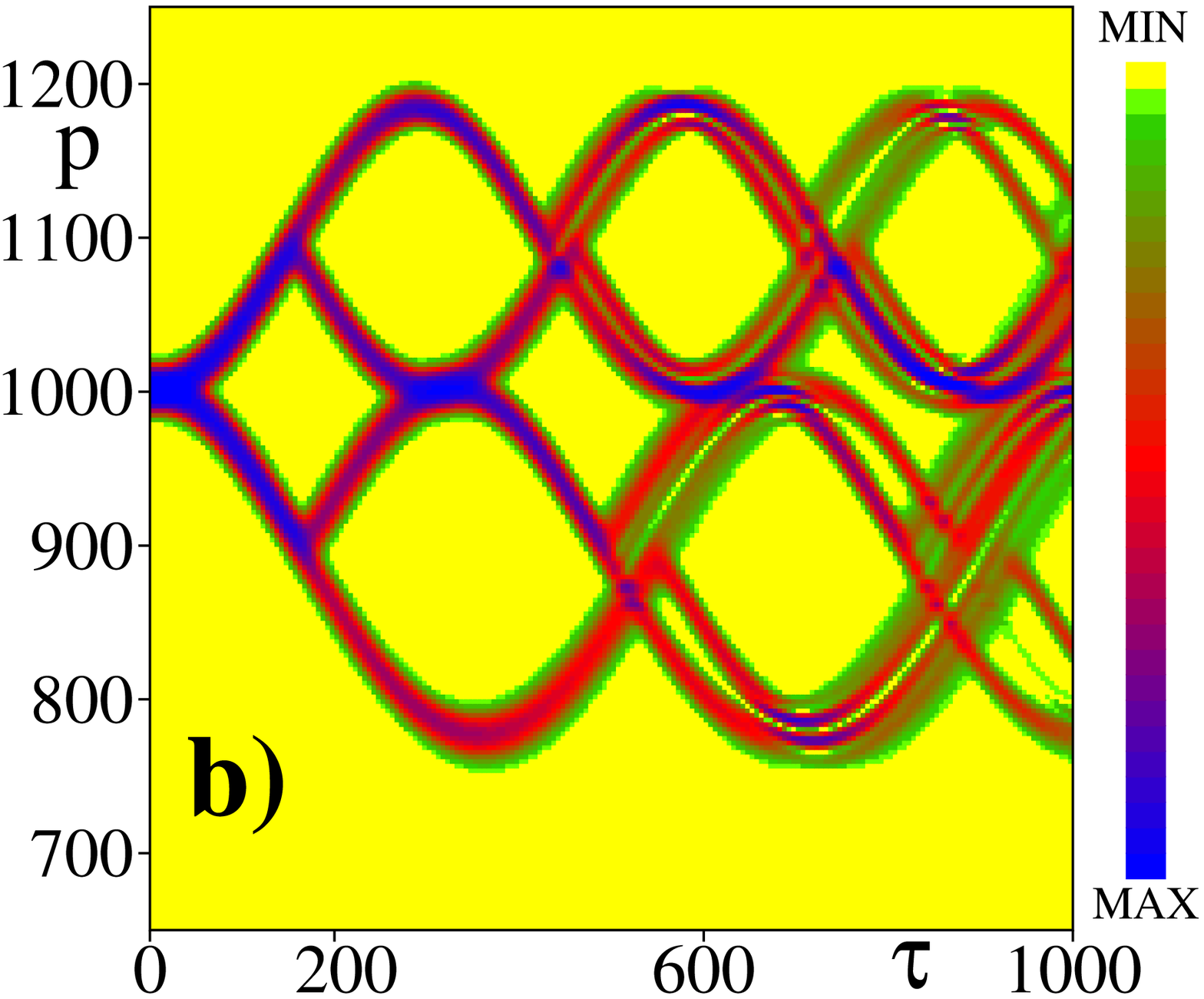}
\caption{(Color online) Momentum probability distribution ${\cal  P}(p,\tau)$ 
of a Gaussian wave packet vs time with $p_0=1000, \sigma_p^2 =50$, 
and $\omega_r=10^{-5}$ at (a) $\Delta=0.3$, adiabatic motion, 
and (b) $\Delta=0.1$, motion 
with nonadiabatic transitions. The color codes the values of 
${\cal  P}(p,\tau)$.}
\label{fig1}
\end{figure}

To study all the regimes of the wave-packet motion, 
we simulate Eqs. (\ref{Schodinger}) at different values of the 
Landau--Zener parameter $\kappa$ (\ref{LZparameter}). 
The normalized Doppler shift $\omega_D$ nearby a node of the standing wave is 
estimated to be $\omega_D \simeq \omega_rp_0=0.01$. If we 
choose, say, $\Delta=0.3$ we get the first case in our
nomenclature,  $\kappa\gg 1$, with exponentially 
small probability of nonadiabatic transitions. The wave
packet, initially prepared in the ground state which is a superposition 
of the dressed states with approximately equal weights,  
splits from the beginning (Fig.~\ref{fig1}a) into two 
components,  $\ket{+}_\Delta$ and $\ket{-}_\Delta$, each of which moves 
in its own nonresonant potential,  $E_\Delta^{(+)}$ or $E_\Delta^{(-)}$. 
We really do not observe in  Fig.~\ref{fig1}a any 
splitting at the nodes, and the motion of the wave packet at $\Delta =0.3$ 
is adiabatic and practically periodic.
If $\kappa \ll 1$ (the motion near the resonance),  one expects to
observe the periodic motion in the two resonant potentials simultaneously  
without any splitting \cite{JETP09}.  

At $\Delta = 0.1$, we get $\kappa \simeq 1$ and expect 
nonadiabatic transitions at the nodes of the standing wave in accordance 
with formula (\ref{LZ}). The initial ground state $\ket{1}$ is now 
a superposition of the dressed states with practically the same weights. 
The initial bifurcation is accompanied by splittings (see Fig.~\ref{fig1}b) 
that can be proved to occur at the nodes of the standing wave. Let us start 
to analyze the wave-packet motion with the $\ket{+}_{\Delta}$-component (the upper 
curve in the figure). The first splitting occurs at $\tau_1^{(+)}\simeq150$. 
It is easy to prove that it is the moment of time when the centroid of the 
$\ket{+}_{\Delta}$-component crosses the first node at $x=\pi/2$: $\tau^{(+)}_1
=\pi/2\omega_r\overline p_{0,1}^{(+)}\simeq150$, where 
$\overline p_{0,1}^{(+)}$ is the average momentum of the centroid of the 
$\ket{+}_{\Delta}$ wave packet between $x=0$ and $x=\pi/2$. Thus, the wave packet, 
crossing the node, splits into two parts. The first one prolongs its motion 
in the potential $E_\Delta^{(+)}$ after passing the point $x=\pi/2$. It is 
the lower curve in Fig.~\ref{fig1}b starting  at $\tau_1^{(+)}\simeq150$. 
The corresponding packet slows down because this component loses its kinetic 
energy going up to the top of the potential $E_\Delta^{(+)}$. 
As to the second trajectory (the upper curve starting 
at $\tau_1^{(+)}$), it appears due to the nonadiabatic transition to the 
potential $E_\Delta^{(-)}$. That is why it accelerates from the beginning 
and reaches its maximal velocity at $x=\pi$. The $\ket{-}_{\Delta}$-component (the 
lower curve starting at $\tau=0$) splits at the first node at $\tau_1^{(-)}
=\pi/2\omega_r\overline p_{0,1}^{(-)}\simeq 156$. In course  of time both the 
components split at every node of the standing wave at the moments 
$\tau_n^{(\pm)}$ that can be estimated with the simple formula
\begin{equation} 
\omega_r\overline p_{n-1,n}^{(\pm)}\tau^{(\pm)}_n=(2n-1)\frac{\pi}{2},
\ n=2,3,\ldots,
\label{split}
\end{equation}
where $\overline p_{n-1,n}^{(\pm)}$ is an average momentum of the  centroid of the corresponding component between the $(n-1)$-th and $n$-th nodes. Such a proliferation at the nodes means a complexification of the atomic wave function both in the  
momentum and position spaces as compared to the adiabatic and resonant cases.

Now we go to the position space and compute the probability  
$|C(x,\tau)|^2 =|C_-(x,\tau)|^2 + |C_+(x,\tau)|^2$  
to be at point  $x$ at time $\tau$. In Fig.~\ref{fig2} we show 
the result of simulation in the case of adiabatic and nonadiabatic 
motion at $\Delta=0.3$ and $\Delta=0.1$   
corresponding, respectively, to Fig.~\ref{fig1}a and b in the momentum space. 
It is a plot of the position probability  
in the frame moving with the initial atomic velocity  
$\omega_rp_0=0.01$ where the slope straight lines mark positions of 
the nodes of the standing wave in the moving frame. At $\Delta=0.3$, 
the evolution is simple 
without any transitions at the nodes (Fig.~\ref{fig2}a). The splitting of the total 
probability $|C(x,\tau)|^2$ is caused by the initial bifurcation  of the wave packet 
due to its bipotential motion.

The situation is cardinally different when we work in the regime  
with nonadiabatic transitions at the field nodes ($\Delta=0.1$). Splitting 
at the nodes in the momentum space (see Fig.\ref{fig1}b) manifest itself 
in the position space in Fig.~\ref{fig2}b. In this case one observes visible 
changes in the proba6bility $|C(x)|^2$ exactly at the node lines.  
It is a clear evidence of the nonadiabatic transitions  
that occur in the specific range of the control parameters, 
$\kappa \simeq 1$. This results in a proliferation  of 
components of the wave packet at the nodes and, therefore, 
a complexification of the wave function both in the momentum and position spaces.

\section{Quantum-classical correspondence and nonadiabatic quantum chaos}

In this section we compare the quantum results, obtained in the preceding 
sections, with those obtained for the same problem but in the semiclassical 
approximation when the translational motion has been treated as a classical one 
\cite{JETPL01,PRA01,JETPL02,JETP03,PRA07}. Coherent semiclassical evolution of a point-like 
two-level atom is governed by the Hamilton-Schr\"odinger equations with the same 
normalization as in the quantum case
\begin{equation}
\begin{gathered}
\dot x=\omega_r p,\quad \dot p=- u\sin x, \quad \dot u=\Delta v,
\\
\dot v=-\Delta u+2 z\cos x, \quad
\dot z=-2 v\cos x,
\end{gathered}
\label{mainsys}
\end{equation}
where 
\begin{equation} 
u\equiv2 \Rre(a_0b_0^*),\ v\equiv-2\Iim(a_0b_0^*),\ z\equiv|a_0|^2-|b_0|^2
\label{uvz}
\end{equation}
are the atomic-dipole components ($u$ and $v$) and population-inversion ($z$), 
and $a_0$ and $b_0$ are the complex-valued probability amplitudes to find the
atom in the excited and ground states, respectively. 
The system (\ref{mainsys}) has two integrals of motion, the total energy
\begin{equation}
H\equiv\frac{\omega_r}{2}p^2-u\cos x-\frac{\Delta}{2}z,
\label{H}
\end{equation}
and the length of the Bloch vector, $u^2+v^2+z^2=1$. 

Equations (\ref{mainsys}) constitute a nonlinear Hamiltonian
autonomous system with two and half degrees of freedom and two 
integrals of motion. It has been shown in Ref.~\cite{PRA07} 
to have positive values of the maximal Lyapunov exponent  $\lambda$ 
in a wide range of values of the control parameters and initial states. 
This fact implies dynamical chaos in the sense
of exponential sensitivity to small changes in initial conditions and/or
control parameters. The result of computation
of the maximal Lyapunov exponent in dependence on
the detuning $\Delta$ and the initial Doppler shift 
$\omega_D= \omega_r p_0$ is shown in
Fig.~\ref{fig3} at $\omega_r=10^{-5}$. In white regions of the plot 
the values of $\lambda$ are almost zero, and the atomic motion is regular 
in the corresponding ranges of $\Delta$ and $\omega_D$. 
In shadowed regions positive values of $\lambda$ imply unstable motion.
At exact resonance, we get $\lambda=0$ 
because at $\Delta=0$ the semiclassical equations of motion (\ref{mainsys}) 
become integrable 
due to an additional integral of motion, $u={\rm const}$. 
We stress that the 
local instability produces chaotic center-of-mass 
motion in a rigid standing wave without any modulation of its 
parameters in difference from the situation with atoms in a periodically 
kicked optical lattice 
\cite{Raizen, Steck01, HH01}. In dependence on the initial conditions 
and the parameter values, an atom 
may oscillate in a well of the lattice or it may have enough kinetic 
energy to overcome the potential barrier. In some cases the center-of-mass 
motion resembles a random walking. 
It means that an atom in a deterministic standing-wave field 
alternates between flying through the lattice, and 
being trapped in its wells. Moreover, it may change the direction of motion in 
a random-like way (see Ref.~\cite{JETP03,PRA07} for coherent Hamiltonian dynamics 
and Refs.~\cite{PRA08,EPL08} for a dissipative one with spontaneous emission 
included).

It follows from (\ref{mainsys}) that the translational motion 
is described by the equation for a nonlinear physical
pendulum with the frequency modulation
\begin{equation}
\ddot x+\omega_r  u(\tau)\sin x=0,
\label{12}
\end{equation}
where $u$ is a function of all the other dynamical variables.
It has been shown in Ref.~\cite{PRA07} that the regime of the center-of-mass  
motion is specified by the character of oscillations of the component 
$u$ of the Bloch vector. In a chaotic regime sudden ``jumps'' of the 
variable $u$ occur when an atom crosses the field nodes.
Figure~\ref{fig4}a demonstrates more or less periodic oscillations 
of $u$ at the detuning value $\Delta=0.3$ at which the corresponding maximal 
Lyapunov exponent is zero (Fig.~\ref{fig3}).
In the chaotic regime at $\Delta=0.1$ $u$ demonstrates shallow oscillations 
interrupted by jumps of different amplitudes upon crossing the nodes (Fig.~\ref{fig4}b).

Approximating the variable $u$ between the nodes by
constant values, the following stochastic map has been constructed 
in Ref.~\cite{PRA07}
\begin{equation}
u_{m}=\sin (\Theta \sin\phi_{m}+\arcsin u_{m-1}),
\label{u_m}
\end{equation}
where $\Theta \equiv \sqrt {\pi \Delta^2/\omega_r p_{\rm{node}}}$ is an 
angular amplitude of the jump,
$u_m$ is a value of $u$ just after the $m$-th node crossing,  
$\phi_m$ are random phases to be chosen in the range $[0,2\pi]$, 
and $p_{\rm{node}}\equiv \sqrt{2H/\omega_r}$
is the value of the atomic momentum at the instant when the atom crosses 
a node (which is the same with a given value of the energy $H$  
for all the nodes).  With given values of $\Delta$,
$\omega_r$ and $p_{\rm{node}}$, the map (\ref{u_m}) has been shown
numerically to give a satisfactory probabilistic distribution of
magnitudes of changes in the variable $u$ just after crossing
the nodes. The stochastic map (\ref{u_m}) is valid under the assumptions
of small detunings ($|\Delta|\ll 1$) and comparatively slow atoms
($|\omega_r p|\ll 1$). Furthermore, it is valid only for those ranges
of the control parameters and initial conditions where the motion
of the basic system (\ref{mainsys}) is unstable. For example,
in those ranges where all the Lyapunov exponents are zero, $u$ becomes
a quasi-periodic function and cannot be approximated by the map 
(see Fig.~\ref{fig4}a).

\begin{figure}[htb]\center
\includegraphics[width=0.45\textwidth,clip]{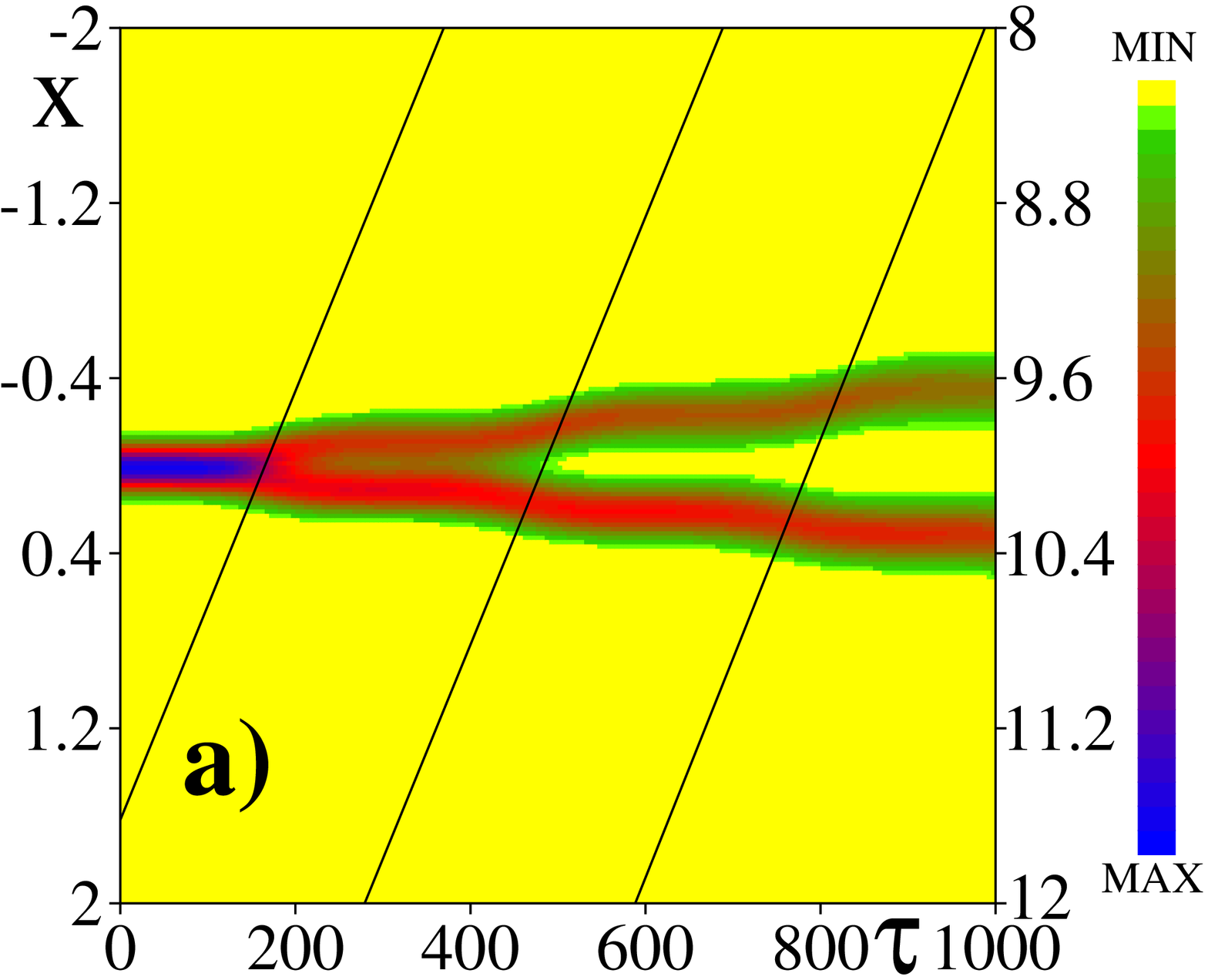}
\includegraphics[width=0.45\textwidth,clip]{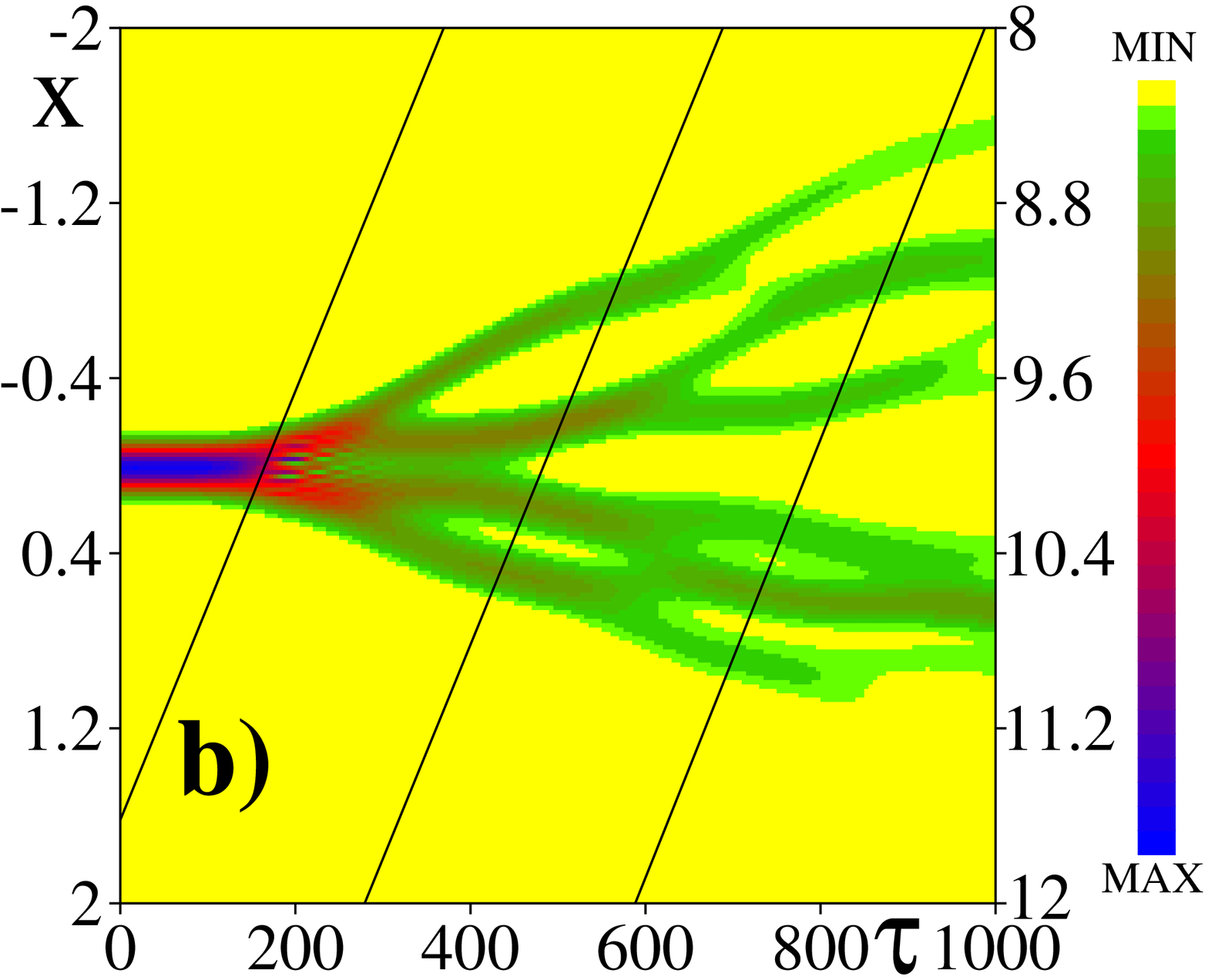}
\caption{(Color online) The position probability $|C(x)|^2$ 
in the moving frame of reference with the slope straight lines marking 
positions of the nodes. (a) Adiabatic motion 
in the position space at $\Delta=0.3$. (b) Wave-packet propagation 
in the position space with nonadiabatic transitions at the field nodes.}
\label{fig2}
\end{figure}

\begin{figure}[htb]\center
\includegraphics[width=0.45\textwidth,clip]{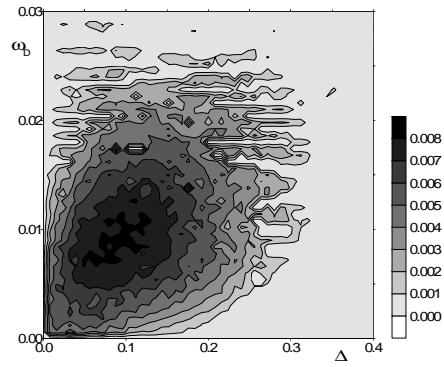}
\caption{Maximal Lyapunov exponent $\lambda$ vs atom-field detuning 
$\Delta$ and the initial Doppler shift $\omega_D= \omega_r p_0$. 
Color codes the values of $\lambda$.}
\label{fig3}
\end{figure}

\begin{figure}[htb]\center
\includegraphics[width=0.45\textwidth,clip]{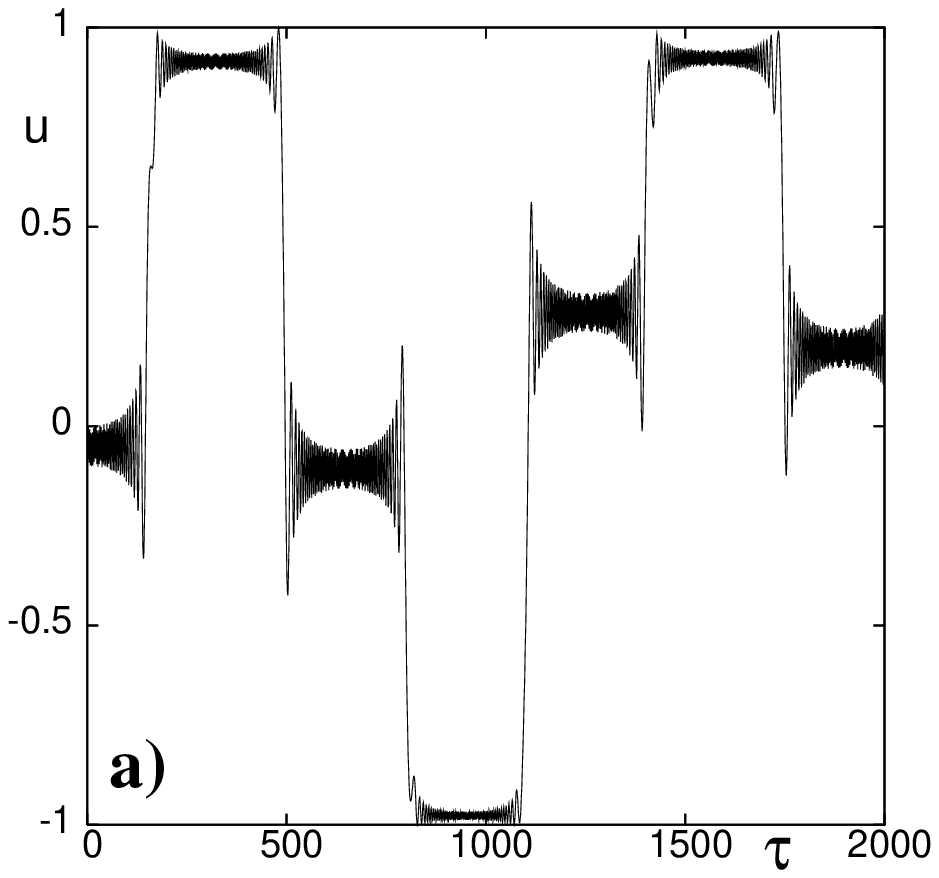}
\includegraphics[width=0.45\textwidth,clip]{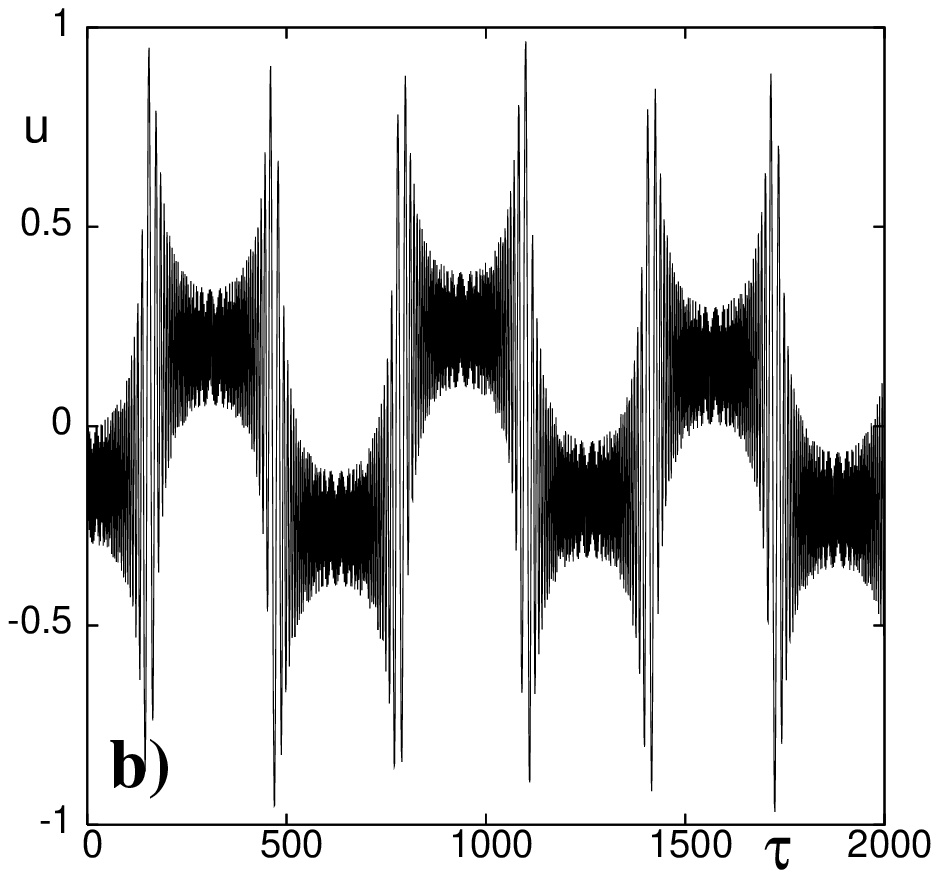}
\caption{Semiclassical evolution of the atomic-dipole component $u$ in (a) 
regular ($\Delta=0.3$) and (b) chaotic ($\Delta=0.1$) regimes of 
the ballistic motion of a point-like atom.}
\label{fig4}
\end{figure}

The key result in the context of the quantum-classical correspondence 
is that the squared angular amplitude of the map (\ref{u_m}) 
is exactly the Landau--Zener parameter (\ref{LZparameter}), i.e., 
$\Theta^2=\kappa $.   
Rewriting the map (\ref{u_m}) for $\arcsin u_{m}$, one gets 
\begin{equation}
\arcsin u_{m}= \sqrt \kappa \sin\phi_{m}+\arcsin u_{m-1},
\label{u_ma}
\end{equation}
where the jump magnitude does not depend on a current value of the variable. 
The map (\ref{u_ma}) visually looks as a random motion of the point along a
circle of unit radius (see Fig.~4 in Ref.~\cite{PRA07}). 
If $\kappa \simeq 1$, then the internal atomic variable $\arcsin u_m$ just 
after crossing the $m$-th node may take with the same probability 
practically any value from the range $[-\pi/2,\pi/2]$. It means semiclassically 
that the momentum of a ballistic atom changes chaotically upon crossing the 
field nodes. In accordance with the quantum formula (\ref{LZparameter}), 
the corresponding atomic wave packet makes nonadiabatic transitions when 
crossing the nodes and splits at each node (see Figs.~\ref{fig1}b and \ref{fig2}b). 
As the result, the wave packet of a single atom 
becomes so complex that it may be called a chaotic one in the sense that 
it is much more complicated than the wave packets propagating adiabatically.
Thus, nonadiabatic wave chaos and semiclassical dynamical chaos occur in 
the same range of the control parameters and are specified by the same 
Landau--Zener parameter $\kappa \simeq 1$. 
In two limit cases with $\kappa \ll 1$ and $\kappa \gg 1$
both the semiclassical and quantized translational ballistic motion 
are regular.

In quantum mechanics there is no well-defined notion of a trajectory in the 
phase space and, hence, the Lyapunov exponents can not be 
computed. In quantum mechanics there is no exponential 
sensitivity to small variations in initial conditions because the time evolution 
of an isolated quantum system is unitary, and the overlap of any two different 
quantum state vectors is a constant in course of time. Moreover, quantum 
phase space is discrete due to the Heisenberg uncertainty principle unlike  
continuous classical phase space.  Namely the continuity of the classical 
phase space provides a possibility of chaotic mixing which exploits more and 
more fine structures in the classical phase space in course of time whereas 
the quantum evolution stops to do that over a rather short Ehrenfest time. 

The semiclassical  (\ref{mainsys}) and quantum 
(\ref{Schodinger}) equations of motion look very different.  The semiclassical  
ones constitute a five-dimensional nonlinear dynamical system of 
ODEs with two integrals of motion that has been shown to be chaotic in a certain 
range of control parameters with exponential sensitivity to small variations  
to initial conditions \cite{PRA07}. 
The quantum ones constitute an infinite-dimensional 
set of linear equations. It is not evident {\it a ~priori} that their solutions 
might demonstrate a kind of correspondence in the same range of the control 
parameters. Nevertheless, a sort of quantum-classical correspondence 
both in regular and chaotic regimes of the center-of-mass  motion 
has been found. It should be stressed that this correspondence manifests itself 
in behavior of the quantum Bloch variable $u$ in the semiclassical  
equations of motion (\ref{mainsys}). 

However, the quantum-classical correspondence is not and could not be absolute
because the Planck constant is equal to 1 with our normalization. 
It cannot tend to 
be zero in order to achieve a classical limit as it could be done with 
an effective Planck constant (see, for example, Refs. \cite{GSZ92,HH01})
depending on the system's parameters. We work in this sense in a deep 
quantum regime. The quantum-classical dualism with cold atoms resembles 
the wave-ray one in a classical wave motion. In the context of this paper 
we might compare ray-like trajectories of atoms with their wave-like motion.  

\begin{figure}[htb]\center
\includegraphics[width=0.45\textwidth,clip]{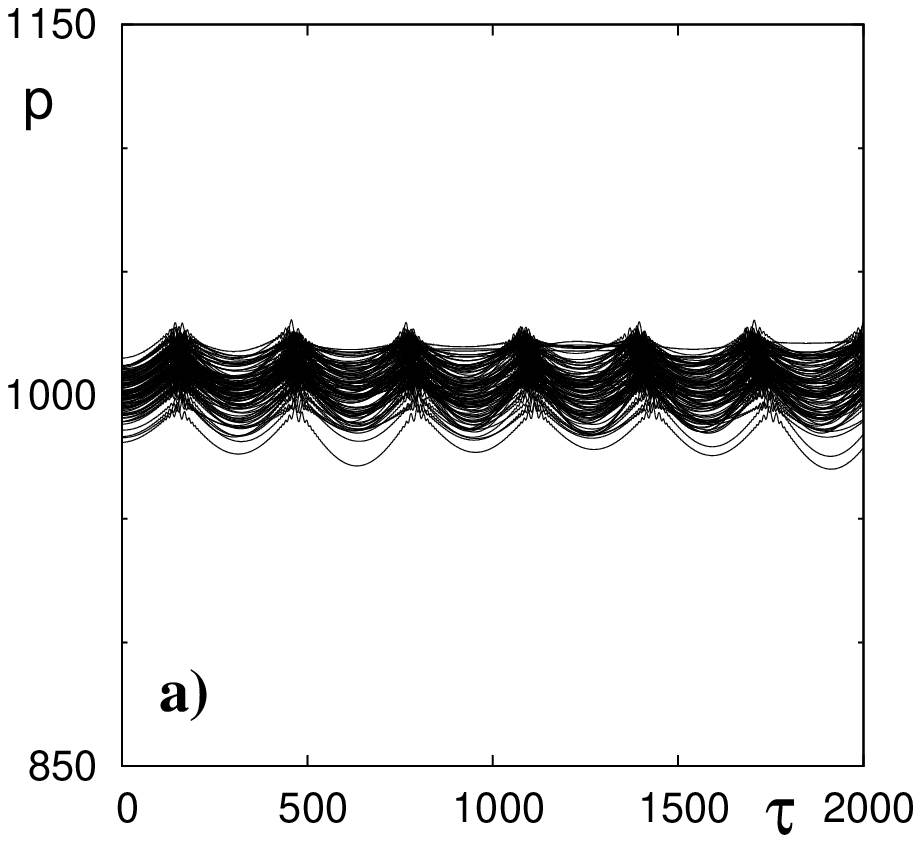}
\includegraphics[width=0.45\textwidth,clip]{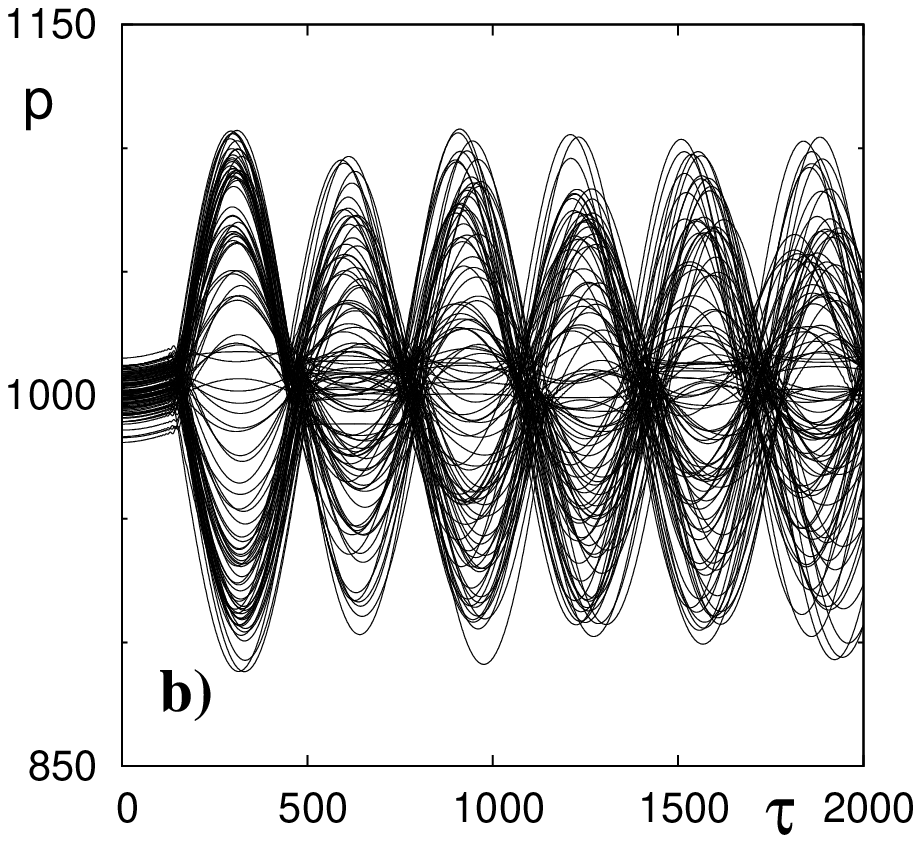}
\caption{Atomic trajectories in the  momentum space computed with the 
classical Gaussian distribution at the same 
parameter's values as in simulation with Gaussian wave packets. 
(a) Regular center-of-mass motion ($\Delta=0.3$) corresponding to 
adiabatic quantum motion in Fig.~\ref{fig1}a and (b) weakly chaotic motion  
($\Delta=0.1$) corresponding to nonadiabatic quantum motion in Fig.~\ref{fig1}b.
}
\label{fig5}
\end{figure}

To illustrate correspondence and difference that inavoidably appears when comparing 
quantum evolution with the classical one (that is only an approximation to 
the quantum one), we compute with Eqs.~(\ref{mainsys}) the evolution of 
a Gaussian distribution over classical momentum 
$p$ and position $x$ with the same parameter's values as in simulation 
of the wave-packet propagation shown in Figs.~\ref{fig1} and 
~\ref{fig2}.  In accordance with the Lyapunov map in Fig.~\ref{fig3}, one expects 
a regular center-of-mass motion at the detuning $\Delta=0.3$ and a weakly chaotic one 
at $\Delta=0.1$. In Figs.~\ref{fig5}a and b evolution of classical momenta 
is shown for the regular ($\Delta=0.3$) 
and chaotic ($\Delta=0.1$) regimes of the center-of-mass motion. Visible 
spreading in $p$  with chaotically moving atoms, as compared to regularly 
moving ones, is one of the signs of classical dynamical chaos. Figure~\ref{fig1} 
demonstrates similar spreading of the momentum probability distribution 
of a Gaussian wave packet with nonadiabatic transitions at $\Delta=0.1$
(Fig.~\ref{fig1}b) as compared to the adiabatically moving wave packet 
at $\Delta=0.3$ (Fig.~\ref{fig1}a). The difference between the classical and 
quantum evolution is also evident: the semiclassical equations of motion 
~(\ref{mainsys}) are not able to simulate the splitting of wave packets 
due to purely quantum effect of motion in two optical potentials simultaneously.

\begin{figure}[htb]\center
\includegraphics[width=0.45\textwidth,clip]{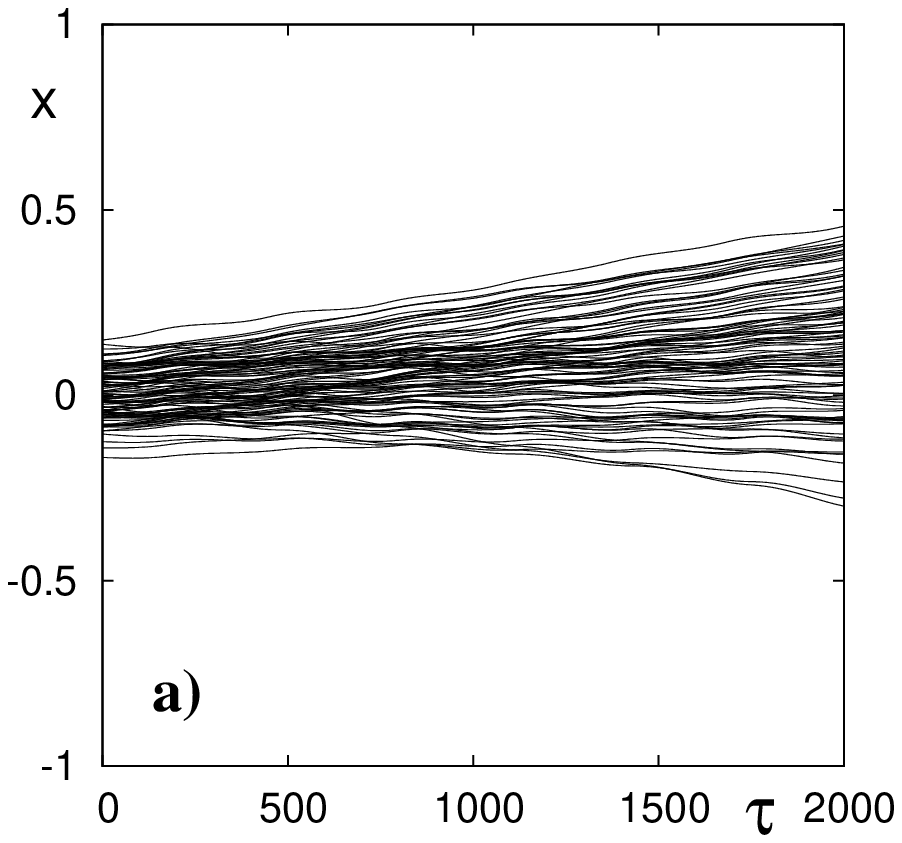}
\includegraphics[width=0.45\textwidth,clip]{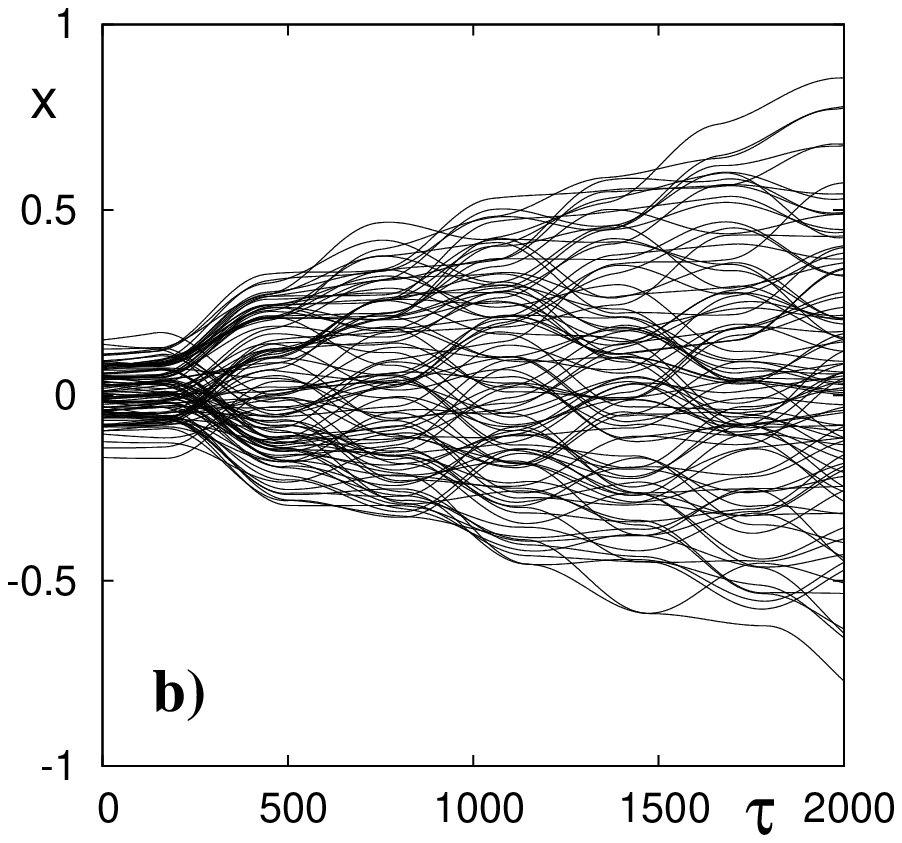}
\caption{Classical atomic trajectories in the moving frame of reference. 
(a) Regular bundle ($\Delta=0.3$) corresponding to 
adiabatic quantum motion in Fig.~\ref{fig2}a and (b) weakly chaotic bundle 
($\Delta=0.1$) corresponding to nonadiabatic quantum motion in Fig.~\ref{fig2}b.}
\label{fig6}
\end{figure}

In Figs.~\ref{fig6}a and b we plot, respectively, regular 
and chaotic trajectories in the frame of reference moving with the initial atomic 
velocity $\omega_rp_0=0.01$. The bundle of chaotically moving atoms 
in Fig.~\ref{fig6}b 
diverges in a short time significantly as compared to the regular one 
in Fig.~\ref{fig6}a. This property can be used to detect chaotic scattering 
in a real experiment with atoms crossing a standing laser wave \cite{JETPL10}. 
As to quantum motion in the position space, it is evident  
that the wave packet with nonadiabatic transitions (Fig.~\ref{fig2}b) 
becomes much broader in course of time due to splitting at the standing-wave 
nodes than the adiabatic wave packet 
in Fig.~\ref{fig2}a resembling a broadening of the bundle of chaotic point-like atoms 
(Fig.~\ref{fig6}b) as compared to the regular one (Fig.~\ref{fig6}a). 
However, we do not observe any splitting of the classical bundles 
because a classical trajectory simulates only 
the motion of the centroid of a quantum wave packet and cannot simulate  
of course its splitting due to purely quantum effect of motion in two 
optical potentials simultaneously. There is only one optical potential in 
the semiclassical approximation.

\section{Conclusion}

We have studied coherent dynamics of ballistic atomic wave  packets in 
a one-dimensional standing-wave laser field. The problem has been considered 
in the momentum representation and in the dressed-state basis where the motion 
of a two-level atom was interpreted as a motion  in two optical potentials. 
The character of that motion has been shown to depend strongly on the value of 
the Landau--Zener parameter $\kappa$ ~(\ref{LZparameter}). 
If $\kappa\gg 1$, then the probability of transitions from one of the 
potential to another one, which is described by the Landau--Zener formula 
(\ref{LZ}), is exponentially small. Under such a condition, atoms move  
in the adiabatic regime. If $\kappa\ll 1$, the formula 
(\ref{LZ}) gives almost unity probability to change the potential when 
crossing the nodes. In the  
intermediate case,  $\kappa\simeq 1$, the probabilities 
for an atom to change or not  to change the nonresonant potential, when 
crossing a node, are of the same order. The corresponding nonadiabatic 
transitions manifest themselves as a splitting of the atomic wave packets in the 
momentum space when their centroids cross the nodes. 
This nonadiabatic quantum chaos
occurs exactly in the same range of the detuning and the Doppler 
shift where the semiclassical dynamics has been shown to be chaotic. 
It is remarkable that the same Landau--Zener parameter $\kappa$ specifies both  
semiclassical and quantum chaos with ballistic atoms in a deterministic 
optical lattice. 

We hope that the results obtained can be used to study manifestations of 
quantum chaos with Bose-Einstein condensates in optical lattices \cite{Morsch} 
with coupled degrees of freedom. From the theoretical point of view, 
the dynamics of condensates of ultracold atoms is described correctly 
by the Gross-Pitaevskii equation which is a kind of a {\it nonlinear} 
Schr\"odinger equation with possible chaotic solutions. 
Experimentally, one of the possibilities is to prepare 
two Bose-Einstein condensates in different internal states \cite{Matt}. 
Another possibility can be realized with a Bose-Einstein condensate in an optical 
lattice subject to a static tilted force \cite{Jona,Kolovsky,Tay}.        
Viewing transitions between the Bloch bands of a condensate in such a tilted 
optical lattice as a two-state problem, we get a mesoscopic quantum system 
with coupled different degrees of freedom (I am thankful to an anonymous 
referee for that comment). 

\section*{Acknowledgments}
This work was supported  by the Russian Foundation for Basic Research
(project no. 09-02-00358), by the Integration grant from the Far-Eastern 
and Siberian branches of the Russian Academy of Sciences, and by the Program
``Fundamental Problems of  Nonlinear Dynamics'' of the Russian
Academy of Sciences. I thank L.E. Konkov and V.O. Vitkovsky for their help 
in preparing some figures.


\end{document}